\documentclass[aps,english,showpacs,twocolumn,pra]{revtex4-1}
\usepackage{amsfonts}
\usepackage{amssymb}
\usepackage{amsmath}
\usepackage{graphicx}
\usepackage{epsfig}
\usepackage{color}

\begin{document}

\title{Non-Hermitian phase transition and eigenstate localization induced by
asymmetric coupling}
\author{P. Wang}
\author{L. Jin}
\email{jinliang@nankai.edu.cn}
\author{Z. Song}
\email{songtc@nankai.edu.cn}
\affiliation{School of Physics, Nankai University, Tianjin 300071, China}

\begin{abstract}
We investigate a uniformly coupled non-Hermitian system with asymmetric
coupling amplitude. The asymmetric coupling equals to a
symmetric coupling threaded by an imaginary gauge field. In a closed
configuration, the imaginary gauge field leads to an imaginary magnetic
flux, which induces a non-Hermitian phase transition. For an open boundary,
the imaginary gauge field results in an eigenstate localization. The
eigenstates under Dirac and biorthogonal norms and the scaling laws are
quantitatively investigated to show the affect of asymmetric coupling
induced one-way amplification. However, the imaginary magnetic flux does not
inevitably induce the non-Hermitian phase transition for systems without
translation invariance, this is elucidated from the non-Hermitian phase
transition in the non-Hermitian ring with a single coupling defect. Our
findings provide insights into the non-Hermitian phase transition and
one-way localization.
\end{abstract}

\maketitle

\section{Introduction}

The parity-time ($\mathcal{PT}$) symmetry confinement enables purely real
spectrum of non-Hermitian system \cite%
{Bender,NM,El-G,SLonghi,Fan,NatPhoton,SKG,Alu2019}. In $\mathcal{PT}$
symmetric non-Hermitian systems, the degree of non-Hermiticity determines a
phase transition \cite{AGuo} at the exceptional points \cite%
{EP,EP2,BZhen,Doppler,HXu,CTChanPRX,Assawaworrarit,HJing2017,JL97,XLCui},
where\ the system is defective and the intensity polynomial increases \cite%
{WP}. Exceptional points are valuable in quantum metrology and sensing \cite%
{NMParaEstimation,JW,YXLiu}. The intensity of initial excitation oscillates
in the exact $\mathcal{PT}$-symmetric phase with purely real eigenvalues
\cite{Makris08,Klaiman,Ruter}. In the broken $\mathcal{PT}$-symmetric phase,
the eigenvalues become conjugation pairs and the intensity exponentially
increases. The non-Hermitian systems exhibit many intriguing dynamical
phenomena due to the nonorthogonality of the eigenmodes. The coherent
perfect absorption \cite{YDChongCPA}, unidirectional
invisibility/reflectionless \cite{Nature2012,LFengNatureMater},
unidirectional perfect absorption \cite{UPA}, unidirectional propagation
\cite{LJin}, and unidirectional lasing \cite{LJin,LYangPNAS,XZhang}. The
nonlinear non-Hermitian systems with asymmetric structures exhibit
intriguing nonreciprocal property \cite{Kominis16}.

The $\mathcal{PT}$-symmetric gain and loss are experimentally realized in
atomic system \cite{Wu}, optical waveguide \cite%
{Ruter,Ruschhaupt,GanainyOL,LFengNatureMater,YFChen16}, resonator \cite%
{BPeng,HJing,He}, and photonic crystal \cite{Cerjan,YFChenPNAS}, electronic
circuit \cite{Kottos}, acoustics systems \cite{Alu,XZhangPRX}. By employing
the gain, optical isolation \cite{LChang}, single mode lasing in $\mathcal{PT%
}$-symmetric coupled resonators \cite{XZhangScience}, unidirectional lasing
due to asymmetric backscattering \cite{PNAS}, topological lasing \cite%
{TopoLasing} have been experimentally demonstrated. Other non-Hermitian
elements include the pure imaginary coupling and the asymmetric coupling.
The pure imaginary coupling $i\gamma $ has been realized in the atomic
systems to demonstrate the anti-$\mathcal{PT}$-symmetry \cite{NatPhysAntiPT}%
. The asymmetric coupling of directional amplification or attenuation \cite%
{Longhi,LFengNC} in coupled resonators, or atomic gases \cite{ZGong}.
Recently, non-Hermitian topological systems have received extensive
attention \cite%
{JL,Zeuner,HZhao,HZhao2,2013,2011,Schomerus,JGong15,Leykam,YXu,HShen,Weimann,Yuce,Lieu,Kunst,KK,Molina,LJLPRB,JHou,Takata,Okugawa,TDas,Torres,VMMA,HZhou,Ezawa,Yoko,Dan,ZZL}%
, in particular, for systems with asymmetric coupling. The asymmetric
coupling induces all the eigenstates localized at the system boundary, which
is called the non-Hermitian skin effect \cite%
{ZWang,CHLee,ZYang,HWang,LJinPRB,RYu,HJiang2019,DasReview}. The skin effect
and edge mode in non-Hermitian system are revealed in using Green's function
method~\cite{Dan}. The non-Hermitian Su-Schrieffer-Heeger models
with stagger asymmetric coupling with and without chiral-inversion symmetry
are investigated in Ref.~\cite{LJinPRB}. The way of non-Hermiticity
appearance is revealed to be critical. The one-way amplification or
attenuation induces the nonzero non-Hermitian Aharonov--Bohm (AB) effect and
the skin effect. The chiral-inversion symmetry protect the bulk-boundary
correspondence in the non-Hermitian topological systems, and prevents the
non-Hermitian skin effect. In non-Hermitian topological systems, the
non-Hermitian AB effect induces the non-Hermitian phase transition and
topological phase transition \cite{ZWang,CHLee,ZYang,HWang,LJinPRB};
however, whether the non-Hermitian AB effect induces non-Hermitian phase
transition in the non-Hermitian system in general cases is an interesting
question.

In this work, we investigate the non-Hermitian phase transition induced by
asymmetric coupling amplitude. The asymmetric coupling effectively generates
a gauge invariant imaginary field; the imaginary gauge field in a closed
area induces an imaginary magnetic flux, which does not inevitably result in
a non-Hermitian phase transition; in particular, for the
non-Hermitian systems that are not translation invariant. This is elucidated
through a non-Hermitian nonuniform ring system with single defective
coupling. The situation dramatically differs from that in the translation
invariant topological systems, where the imaginary gauge field enters the
wave vector. The momentum becomes a complex number and induces the
non-Hermitian phase transition. The non-Hermitian skin effect found in
topological systems has renewed the research interest for the localization
effect. We consider a uniform chain to quantitatively investigate the
non-Hermitian localization without the influence of system topology. We
concentrate on capturing the properties of asymmetric coupling induced
eigenstates localization; alternatively, the localization is a
unidirectional amplification. Therefore, the averaged inverse participation
ratio (IPR) under the definition of Dirac norm is investigated. At weak
non-Hermiticity, the averaged IPR is inversely proportional to the system
size when system size is small; at large non-Hermiticity, the eigenstates
are tightly localized at the system boundary and the IPR is insensitive to
the system size. In contrast, the gauge invariant field does not affect the
biorthogonal norm of the eigenstates; the biorthogonal IPR is inversely
proportional to the system size as the extended states although the Dirac
probability distributions are localized. Our findings provide insights for
the non-Hermitian asymmetric coupling, non-Hermitian skin effect, and
non-Hermitian phase transition.

The paper is organized as follows. In Sec. II, we introduce the
non-Hermitian system with asymmetric coupling. In Sec. III, we exactly solve
the system and show its energy spectrum. In Sec. IV, we study a nonuniform
case with single coupling difference. In Sec. V, the eigenstates of system
under open boundary condition are investigated, the scaling law of the
eigenstates is presented. In Sec. VI, the biorthogonal IPR of the
eigenstates is discussed. Our results are summarized in Sec. VII.

\section{Model}

\begin{figure}[tb]
\includegraphics[ bb=0 0 520 480, width=8.7 cm, clip]{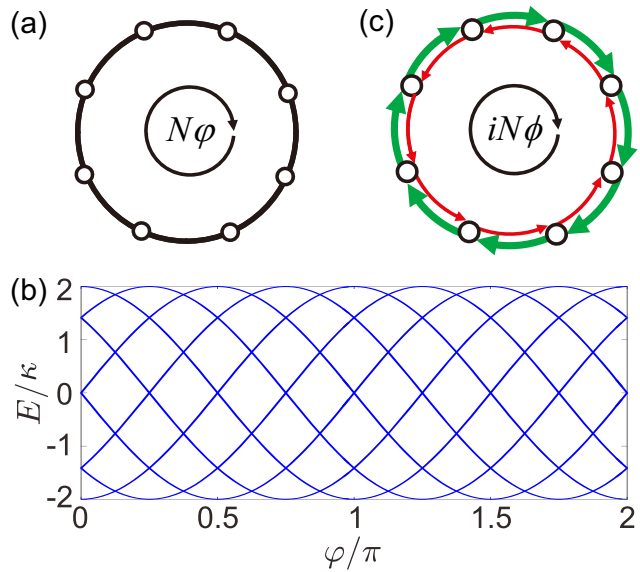}
\caption{(a) Schematic of the uniform ring enclosed with magnetic flux $N\protect\varphi $. (b) Energy spectrum of a uniform ring under the influence
of magnetic flux. The system size is chosen $N=8$ and the parameter is $\protect\kappa =1$ in the plot. (c) Schematic of the uniform ring with
asymmetric coupling strength. The green (red) arrow indicates the tunneling
in the clockwise (counterclockwise) direction. The asymmetric coupling
mimics a gauge field, which induces the imaginary magnetic flux $-iN\protect\phi $.}
\label{fig1}
\end{figure}

Before introducing the non-Hermitian system, we first consider a discrete
one-dimensional tight-binding lattice system with uniform coupling strength.
The coupling amplitude has a Peierls phase factor $e^{i\varphi }$ in the
front, the system Hamiltonian reads%
\begin{equation}
H_{0}=\kappa \sum_{j=1}^{N}(e^{i\varphi }a_{j}^{\dagger }a_{j+1}+\mathrm{H.c.%
}),
\end{equation}%
where $a_{j}^{\dagger }$ ($a_{j}$) is the creation (annihilation) operator
of site $j$. The tight-binding system is schematically illustrated in Fig. %
\ref{fig1}(a) in a closed ring configuration under periodical boundary
condition $a_{j}=a_{j+N}$ and $a_{j}^{\dagger }=a_{j+N}^{\dagger }$ for
total system size $N$. The Peierls phase factor $e^{i\varphi }$ in the
couplings induces a magnetic flux $N\varphi $ in the $N$-site ring system.
The dispersion relation is given by
\begin{equation}
E=2\kappa \cos \left( k+\varphi \right) ,
\end{equation}%
where the wave vector $k$ in the momentum space is discrete, being $k=2n\pi
/N$ with integer $n=1,\cdots ,N$. The corresponding eigenstate for $E$ is a
plane wave with momentum $k$ in the form of $\left\vert \psi
_{k}\right\rangle =\sqrt{1/N}\sum_{j=1}^{N}e^{ikj}a_{j}^{\dagger }\left\vert
\mathrm{vac}\right\rangle $. Under the influence of the magnetic flux $%
N\varphi $, the momentum effectively shifts by a value $\varphi $; however,
the dispersion relation remains unchanged. The energy spectrum of the
uniform ring under the influence of magnetic flux is depicted in Fig. \ref%
{fig1}(b) for an $N=8$ system. Eight energy levels are all changed to their
opposite energies after their momentum shifted by $\pi $ at $\varphi =\pi $.

In Fig. \ref{fig1}(c), the uniform tight-binding lattice with non-Hermitian
asymmetric coupling is shown. The asymmetric coupling strength indicates the
different amplitudes when particles or photons tunneling in opposite
directions. The Hamiltonian reads $H=\sum_{j=1}^{N}(\alpha a_{j}^{\dagger
}a_{j+1}+\beta a_{j+1}^{\dagger }a_{j})$. Considering real $\alpha ,\beta >0$
without loss of generality, the coupling amplitudes can be rewritten as $%
\alpha =\sqrt{\alpha \beta }\sqrt{\alpha /\beta }$ and $\beta =\sqrt{\alpha
\beta }\sqrt{\beta /\alpha }$. The Hamiltonian reduces into another form of%
\begin{equation}
H=\sqrt{\alpha \beta }\sum_{j=1}^{N}(e^{\phi }a_{j}^{\dagger
}a_{j+1}+e^{-\phi }a_{j+1}^{\dagger }a_{j}),
\end{equation}%
where $e^{\phi }=\sqrt{\alpha /\beta }$; the Hamiltonian $H$ has a uniform
symmetric coupling, but threaded by an imaginary magnetic flux \cite{LJinPRB}%
. The nonreciprocal amplification or attenuation factor accumulates when
particles or photons circling along one direction in the ring configuration;
the factor accumulated after circling one round gives $e^{\pm N\phi }$ in
opposite direction. The accumulation can be understood from the view point
of imaginary gauge field and imaginary magnetic flux, which are both gauge
invariant under gauge transformations \cite{Longhi,LFengNC}. The
coupled ring resonator array is a physical realization of $H$ \cite{Longhi},
where the main resonators are evanescently coupled through auxiliary
resonators. To introduce effective asymmetric coupling between the main
resonators, the auxiliary ring resonators have half perimeter gain and half
perimeter loss. Photons tunneling between the main resonators are thus
direction dependent, amplified in one direction and attenuated in the
opposite direction; this results in the asymmetric coupling in $H$. In the
coupled resonator array, the light intensity distribution of eigenstate
reflects the localization effect.

Set $\bar{b}_{j}=e^{-j\phi }a_{j}^{\dagger }$, $b_{j}=e^{j\phi }a_{j}$, the
operators satisfy the commutation relation $[b_{l},\bar{b}%
_{j}]=[a_{l},a_{j}^{\dagger }]=\delta _{lj}$. We obtain a uniform ring
Hamiltonian with symmetric coupling strength $\sqrt{\alpha \beta }$\ except
one asymmetric coupling at the connection between the head and tail. The
only asymmetric coupling is $\sqrt{\alpha \beta }e^{\pm N\phi }$\ associated
with the amplification or attenuation factor $e^{\pm N\phi }$. The
non-Hermitian Hamiltonian is in the form of $H=\sqrt{\alpha \beta }%
[\sum_{j=1}^{N-1}(\bar{b}_{j}b_{j+1}+\bar{b}_{j+1}b_{j})+e^{i\Phi }\bar{b}%
_{N}b_{1}+e^{-i\Phi }\bar{b}_{1}b_{N}]$, where the amplification or
attenuation factor is given by $e^{i\Phi }=(\sqrt{\alpha /\beta })^{N}$,
which yields an effective imaginary magnetic flux $\Phi =-iN\phi $\ threaded
in the ring configuration. A particle or photon circling one round in the
ring accumulates an additional amplification (attenuation) factor $e^{i\Phi }
$\ ($e^{-i\Phi }$) on its wavefunction in the clockwise (counterclockwise)
directions.

\section{Energy spectrum}

The eigenvalues of the non-Hermitian system are obtained after applying a
Fourier transformation
\begin{equation}
a_{k}^{\dagger }=\frac{1}{\sqrt{N}}\sum_{j=1}^{N}e^{ikj}a_{j}^{\dagger },
\end{equation}%
where $k=2\pi n/N$ with integer $n=1,\cdots ,N$. The Hamiltonian $H$ in the
momentum space is given by%
\begin{equation}
H=\sqrt{\alpha \beta }\sum_{k}[e^{ik+\phi }+e^{-\left( ik+\phi \right)
}]a_{k}^{\dagger }a_{k}.
\end{equation}%
Then, the eigenvalues of $H$ are given by%
\begin{equation}
E=2\sqrt{\alpha \beta }\cos \left( k-i\phi \right) .
\end{equation}%
The energy spectrum of the non-Hermitian lattice with an imaginary magnetic
flux has an identical form of expression in comparison with the Hermitian
system threaded by a real magnetic flux. In contrast to the real magnetic
field that shifts the wave vector $k$ in the momentum space, the imaginary
magnetic field results in the wave vector extended into the complex plane.
This induces a non-Hermitian phase transition. The Aharonov-Bohm effect
occurs only if the gauge field is enclosed in the ring. In an open chain,
the eigenvalues are $E=2\sqrt{\alpha \beta }\cos [n\pi /(N+1)]$, not
affected by the imaginary gauge field.

\begin{figure}[tb]
\includegraphics[ bb=0 0 480 680, width=8.7 cm, clip]{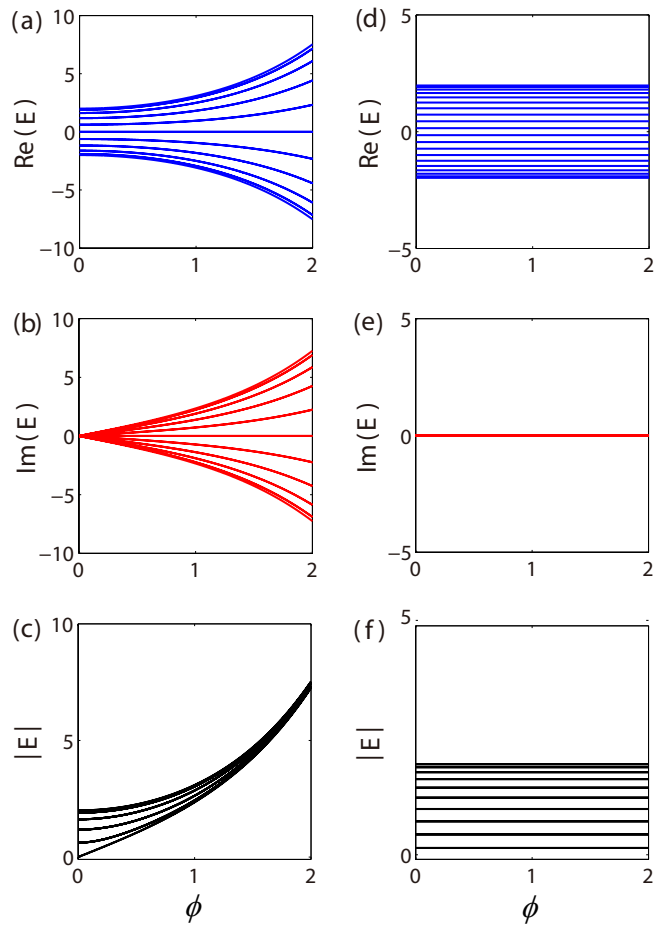}
\caption{Complex energy spectra of the unform coupled non-Hermitian lattice
with asymmetric coupling. (a) Real, (b) imaginary, and (c) absolute values
of the spectrum in the closed ring configuration. (d) Real, (e) imaginary,
and (f) absolute values of the spectrum in the open chain configuration. The
system size is $N=20$, $\protect\alpha \protect\beta =1$.} \label{fig2}
\end{figure}

The energy spectrum of the non-Hermitian lattice [Fig. \ref{fig1}(c)] is
depicted in Fig. \ref{fig2} as a function of the degree of coupling
asymmetry. Under periodical boundary condition, the asymmetry in the
coupling ($\alpha \neq \beta $) leads to a complex spectrum in Figs. \ref%
{fig2}(a)-\ref{fig2}(c) (except the momentum $k=\pi ,2\pi $). The imaginary
part of energy spectrum is enlarged as the degree of coupling asymmetry
(imaginary magnetic flux) increases. Under open boundary condition, the
degree of asymmetry in the coupling does not influence the spectrum and the
spectrum is entirely real.

\section{Nonuniform system}

The non-Hermitian asymmetric coupling does not inevitably induce the
non-Hermitian phase transition. In this section, we discuss a uniform
non-Hermitian system with single different coupling between the head and
tail of the periodical system. The Hamiltonian reads%
\begin{eqnarray}
H_{\mathrm{Non}} &=&\sqrt{\alpha \beta }[\sum_{j=1}^{N-1}(e^{\phi
}a_{j}^{\dagger }a_{j+1}+e^{-\phi }a_{j+1}^{\dagger }a_{j})  \notag \\
&&+Je^{\phi }a_{N}^{\dagger }a_{1}+Je^{-\phi }a_{1}^{\dagger }a_{N}],
\label{NS}
\end{eqnarray}%
where $e^{\phi }=\sqrt{\alpha /\beta }$, $J$ is an enlarged/diminished
amplitude, indicating the strength difference between the normal uniform
coupling and the abnormal coupling. Considering $H_{\mathrm{Non}}$ in a
gauge field, the energy spectrum of $H=\sqrt{\alpha \beta }%
[\sum_{j=1}^{N-1}\left( \bar{b}_{j}b_{j+1}+\bar{b}_{j+1}b_{j}\right)
+Je^{N\phi }\bar{b}_{N}b_{1}+Je^{-N\phi }\bar{b}_{1}b_{N}]$ is studied. The
dispersion relation still in the form of $E_{n}=2\sqrt{\alpha \beta }\cos
k_{n}$. From the Schr\"{o}dinger equations, we obtain the critical equation
for the momentum $k_{n}$, which is a transcendental equation in the form of

\begin{equation}
2J\cosh \left( \phi N\right) \sin k_{n}=\sin \left[ k_{n}\left( 1+N\right) %
\right] +J^{2}\sin \left[ k_{n}\left( 1-N\right) \right] .  \label{TS}
\end{equation}%
We set $k_{n}=2n\pi /N+\theta _{n}$. For a large $N$, $N\theta _{n}$ is
nonzero finite and $\theta _{n}\approx 0$. Equation (\ref{TS}) reduces to%
\begin{equation}
\sin \left( \eta _{n}+\theta _{n}N\right) =2J\cosh \left( \phi N\right) \sin
\left( 2n\pi /N\right) /\xi _{n},
\end{equation}%
(except for $n=N/2$, $N$), where $\eta _{n}$ satisfies%
\begin{equation}
\left\{
\begin{array}{c}
\sin \eta _{n}=\left( 1+J^{2}\right) \sin \left( 2n\pi /N\right) /\xi _{n}
\\
\cos \eta _{n}=\left( 1-J^{2}\right) \cos \left( 2n\pi /N\right) /\xi _{n}%
\end{array}%
\right. ,
\end{equation}%
and%
\begin{equation}
\xi _{n}=\sqrt{\left( 1-J^{2}\right) ^{2}+4J^{2}\sin ^{2}\left( 2n\pi
/N\right) }.
\end{equation}

Therefore, the condition of spectrum with complex energy level is the
occurrence of complex $\theta _{n}$; which indicates the non-Hermitian phase
transition \cite{El-G} and requires $\left\vert \sin \left( \eta _{n}+\theta
_{n}N\right) \right\vert >1$. The condition for the energy level $n$ to be
complex is given by%
\begin{equation}
\sinh ^{2}\left( \phi N\right) >\frac{\left( 1-J^{2}\right) ^{2}}{4J^{2}\sin
^{2}\left( 2n\pi /N\right) },
\end{equation}%
the most fragile energy level is the one with $2n\pi /N$ that most close to $%
\pi /2$. For fixed $\phi $, taking system size $N=4n$ as an illustration,
where $2n\pi /N=\pi /2$ at $n=N/4$ and the critical coupling amplitude $%
J_{c} $ for exceptional points~\cite{EP2} is%
\begin{equation}
J_{c}=\pm e^{\pm \phi N}.  \label{threshold}
\end{equation}

\begin{figure}[tb]
\includegraphics[ bb=17 109 573 674, width=8.7 cm, clip]{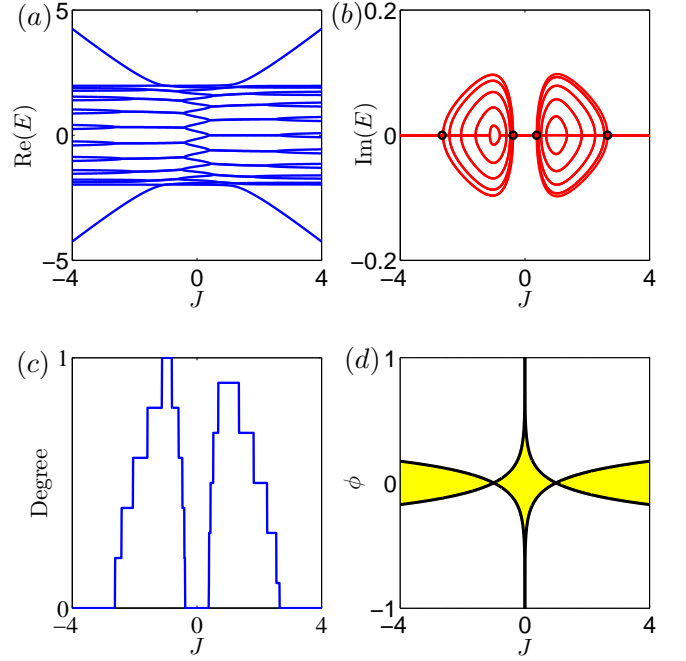}
\caption{(a) Real and (b) imaginary parts of energy spectrum of a nonuniform
non-Hermitian system with single different coupling. The black circles are $J_{c}$ from Eq. (\protect\ref{threshold}). (c) Degree of complex energy
levels in the spectrum. (d) Phase diagram on the parameter $J$-$\protect\phi
$ plane. The yellow area indicates the entirely real spectrum, while the white area indicates the complex spectrum. The system size is $N=20$, and the parameter is $\protect\sqrt{\protect\alpha /\protect\beta }=1.05$
in (a), (b), and (c). } \label{fig3}
\end{figure}

The real and imaginary parts of the energy levels at the parameter $\sqrt{%
\alpha /\beta }=1.05$ are depicted in Figs.~\ref{fig3}(a) and~\ref{fig3}(b).
In Fig. \ref{fig3}(a), as $J$ increases, two energy levels are out of the
energy bands, forming the bound states. In Fig. \ref{fig3}(b), we notice
four critical $J_{c}$, the regions with zero imaginary part inside the
central two $J_{c}$ and outside the outer two $J_{c}$ are phase with
entirely real spectrum. At different $J$, the number of complex energy
levels differs, which can be seen in Fig. \ref{fig3}(b). Considering Eq. (%
\ref{TS}) in the case of odd $N$, the spectrum of system with $-J$ possesses
$\pi $ shift of momentum ($k\rightarrow k+\pi $) for system with $J$; thus,
the spectrum of system with $-J$ is inversed ($E_{k}\rightarrow -E_{k}$) in
comparison to system with $J$. This conclusion is not valid for even $N$.
The number of complex energy levels differs at $J=\pm 1$, this attributes to
the energy level $n=N$, $N/2$. In the case of $n=N$ or $N/2$, Eq. (\ref{TS})
reduces to%
\begin{eqnarray}
2J\cosh \left( \phi N\right) \sin \theta _{n} &=&\left( 1+J^{2}\right) \sin
\theta _{n}\cos \left( \theta _{n}N\right)  \notag \\
&&+\left( 1-J^{2}\right) \cos \theta _{n}\sin \left( \theta _{n}N\right) .
\label{TS2}
\end{eqnarray}%
For an odd $N$ and $n=N$, Eq.~(\ref{TS2}) reduces to $\cos \left( \theta
_{n}N\right) =\cosh \left( \phi N\right) $ at $J=1$, obviously, $\theta
_{n}=i\phi $; at $J=-1$, we have $\theta _{n}=\pi +i\phi $. This indicates
that one real energy level exists for odd $N$ when $\left\vert J\right\vert
=1$. For even $N$ and $n=N/2,N$, $\theta _{n}=i\phi $ for $J=1$ only. For $%
J=-1$, Eq.~(\ref{TS2}) reduces to $\cos \left( \theta _{n}N\right) =-\cosh
\left( \phi N\right) $, one can rewrite it as%
\begin{equation}
\cos \left[ N\theta _{n}+\left( N-1\right) \pi \right] =\cosh \left( \phi
N\right) ,
\end{equation}%
we can obtain $\theta _{n}=i\phi +\left( N-1\right) \pi /N$. Thus, all the
energy levels are complex at $J=-1$, and two real energy levels exist at $%
J=1 $ [Fig. \ref{fig3}(b)].

We define the number of complex energy levels over the total number of
energy levels as the degree of complex energy levels \cite{Degree}, which is
depicted as a function of $J$ in Fig. \ref{fig3}(c). The degree of complex
energy levels is depicted as the signature of plateau, this is a result of
the finite size effect. The plateau turns steep when $J$ approaches to $%
J_{c} $ in the middle because the imaginary parts of energy levels become
sensitive as depicted in Fig.~\ref{fig3}(b). The phase diagram for
non-Hermitian phase transition is depicted in Fig.~\ref{fig3}(d). The yellow
areas are regions with entirely real spectrum; the white areas are regions
with complex energy levels. The non-Hermitian phase transition occurs at the
black curves. At $J=4$ in the yellow region, the system lies in
the phase with entirely real spectrum in spite of the nonzero imaginary
magnetic flux is present. At $J=1$, any nonzero imaginary magnetic flux
leads to the non-Hermitian phase transition. This reflects that the
imaginary magnetic flux does not inevitably induce the non-Hermitian phase
transition for the nonuniform lattice with a single coupling defect.

\section{The Eigenstates in open system}

\begin{figure}[tb]
\includegraphics[ bb=30 110 647 680, width=8.7 cm, clip]{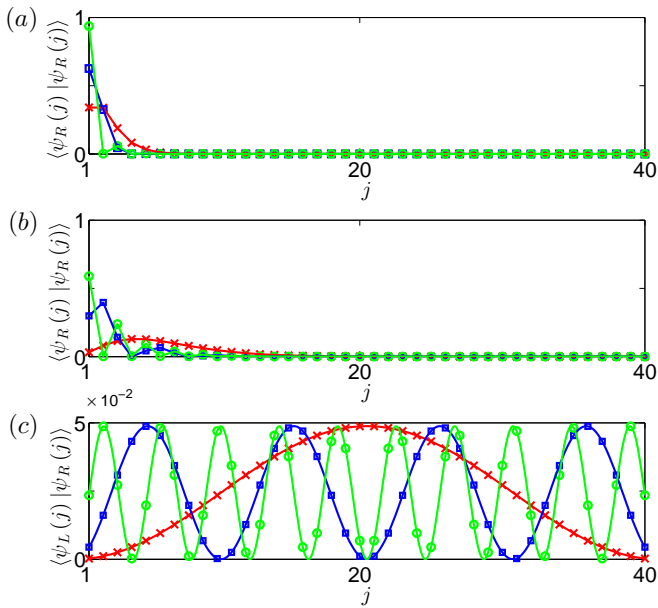}
\caption{The eigenstates of the non-Hermitian lattice with asymmetric
coupling. We select three eigenstates in the open chain system at a certain
asymmetric coupling. The eigenstates are localized at the lattice boundary
in (a), (b), and are extended in (c). The parameter is $\protect\sqrt{\protect\alpha /\protect\beta }=2$ in (a), (c); and $\protect\sqrt{\protect\alpha /\protect\beta }=5/4$ in (b).}
\label{fig4}
\end{figure}

The imaginary gauge field does not influence the system energy spectrum
under open boundary condition at $J=0$ of Hamiltonian Eq. (\ref{NS}), where
the spectrum is entirely real. Although the eigen spectrum is unchanged
under open boundary condition as depicted in Figs. \ref{fig2}(d)-\ref{fig2}%
(f), the imaginary gauge field induces all the eigenstates
localized at one system boundary. The scaling law of the localization effect
only depends on the strength of the imaginary gauge field. The localization
is a one-way amplification that robust to disorder.

Under open boundary condition, the Hamiltonian reads
\begin{equation}
H_{\mathrm{chain}}=\sqrt{\alpha \beta }\sum_{j=1}^{N-1}(e^{\phi
}a_{j}^{\dagger }a_{j+1}+e^{-\phi }a_{j+1}^{\dagger }a_{j}).
\end{equation}%
The coupling between the head and tail is absent.

We assume the eigenstates as
\begin{equation}
\left\vert \psi _{n}^{\mathrm{R}}\right\rangle =\frac{1}{\Omega _{n}}%
\sum_{j=1}^{N}f_{j}^{n}a_{j}^{\dagger }\left\vert \mathrm{vac}\right\rangle ,
\label{ES}
\end{equation}%
the amplitude $f_{j}^{n}$\ of the eigenstate is in the form of%
\begin{equation}
f_{j}^{n}=e^{-\phi j}\sin \left( k_{n}j\right) ,(j=1,2,...N),
\end{equation}%
and the renormalization factor is%
\begin{equation}
\Omega _{n}=\sqrt{\sum_{j=1}^{N}e^{-2j\phi }\sin ^{2}\left( k_{n}j\right) }.
\end{equation}%
The detailed calculation is in the Appendix. The Dirac norm of
the eigenstates is defined as $\langle \psi _{n}^{\mathrm{R}}\left\vert \psi
_{n}^{\mathrm{R}}\right\rangle $, notice that $\langle \psi _{n}^{\mathrm{R}%
}\left\vert \psi _{n}^{\mathrm{R}}\right\rangle =1$ for the right eigenstate
$\left\vert \psi _{n}^{\mathrm{R}}\right\rangle $ with the renormalization
factor $\Omega _{n}$. The eigenstates are not orthogonal in the
non-Hermitian system \cite{Makris08}; thus the time evolution is nonunitary.
The scalar light wave function in optical system\textbf{\ }under the
paraxial approximation{\ analogies the Schr\"{o}inger equation\ that
characterizing the particle dynamics \cite{LonghiLPR}. The light intensity
is a physical observable, being the Dirac norm of wave. The light intensity
oscillates in the exact }$\mathcal{PT}$\textbf{-}symmetric phase, known as
the power oscillation\textbf{\ }\cite{Ruter}.

In Figs.~\ref{fig4}(a) and \ref{fig4}(b), we depict the Dirac norm
probability distribution of\ typical eigenstates in a system with size $N=40$%
, all the eigenstates are localized at one boundary of the non-Hermitian
lattice, known as the non-Hermitian skin effect \cite%
{ZWang,CHLee,ZYang,HWang,LJinPRB,RYu,HJiang2019,DasReview}. The skin effect
is because of the directional amplification or attenuation of the asymmetric
coupling, thus eigenstate amplitude is one-way amplified ($\alpha >\beta $).

\begin{figure}[tb]
\includegraphics[ bb=25 125 555 650, width=8.7 cm, clip]{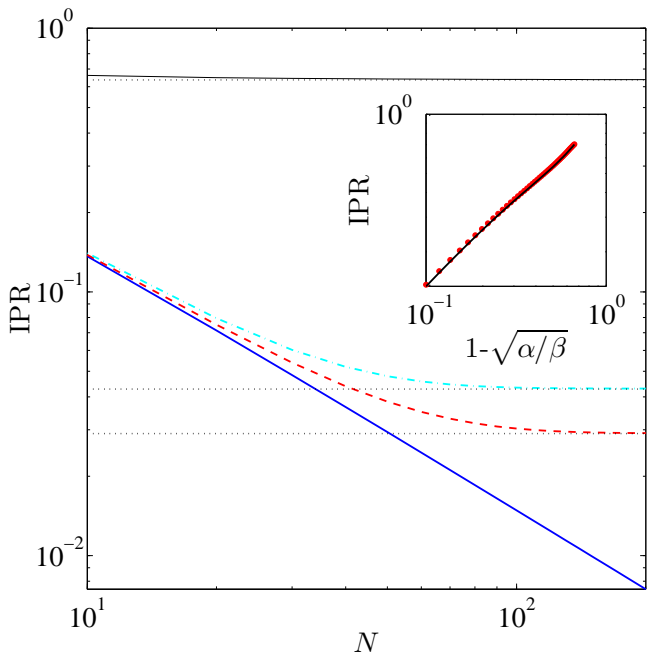}
\caption{The averaged IPR of the eigenstates. The curves are for different
degrees of asymmetry. The solid black $\protect\sqrt{\protect\alpha /\protect\beta }=2.5$, the dash-dotted green $\protect\sqrt{\protect\alpha /\protect\beta }=1.03$, the dashed red $\protect\sqrt{\protect\alpha /\protect\beta }=1.02$, and the dotted blue curve is for the uniform ring with symmetric
coupling as comparison. The asymptotic value at large $N$ is indicated by
the dotted black lines. Insert: The averaged IPR as a function of coupling
asymmetry at $N=40$. The red points are obtained through numerical
simulation and the black lines are plotted from Eq. (\protect\ref{IPRlimit}). }
\label{fig5}
\end{figure}

To investigate the localization effect induced by the asymmetric coupling,
the inverse participation ratio (IPR) is studied. The IPR for eigenstate $%
\left\vert \psi _{n}^{\mathrm{R}}\right\rangle $ is defined as
\begin{equation}
\chi _{n}=\frac{\sum_{j=1}^{N}\left\vert f_{j}^{n}\right\vert ^{4}}{\left(
\sum_{j=1}^{N}\left\vert f_{j}^{n}\right\vert ^{2}\right) ^{2}}.
\end{equation}%
The IPR is valuable in capturing the localization properties of eigenstates.
The IPR of the localization states is insensitive to the system size; in
contrast, the IPR of the extended states is inversely proportional to the
system size. We employ the averaged IPR to reveal the localization
properties of the eigenstates, $\chi =\left( 1/N\right) \sum_{n=1}^{N}\chi
_{n}$, which is
\begin{equation}
\chi =\frac{1}{N}\sum_{n=1}^{N}\frac{1}{\Omega _{n}^{4}}\sum_{j=1}^{N}e^{-4%
\phi j}\sin ^{4}\left( k_{n}j\right) .
\end{equation}%
We can obtain the asymptotic value of the averaged IPR at the limit\textbf{\
}$N\rightarrow \infty $ in the form of%
\begin{equation}
\chi _{c}=\frac{1}{4}\tanh \phi \left( \mathrm{sech}^{4}\phi \tanh \phi
-8\tanh \phi +6\tanh ^{2}\phi +6\right) ,  \label{IPRlimit}
\end{equation}%
which only depends on the asymmetric coupling strength.

In Fig. \ref{fig5}, we depict the averaged IPR as a function of the system
size and asymmetric coupling strength. The asymptotic values of the averaged
IPR [Eq. (\ref{IPRlimit})] at different asymmetric couplings are indicated
by the dotted black lines. The scaling behavior of average IPR reflect the
localization effect of the eigenstates. At small system size, the slope of
IPR approaches $-1$; the exponential decay of the eigenstates and the
localization effect are not obvious to be observed at weak coupling
asymmetry. Therefore, the eigenstates exhibit extended behavior, being
inversely proportional to the system size. At large system size or at large
coupling asymmetry, the slope of IPR in Fig. \ref{fig5} goes to zero; the
localization effect is obvious and the averaged IPR becomes insensitive to
the system size. At significantly strong asymmetry, the eigenstates are
tightly localized at several sites near the system boundary (solid black
line). Insert of Fig. \ref{fig5} depicts the relation between the asymptotic
values of the averaged IPR and the asymmetric coupling; the system size is
chosen $N=40$; $\chi _{c}$ is approximately proportional to $1-\sqrt{\alpha
/\beta }$. The localization length of the eigenstates is $[\ln (\sqrt{\alpha
/\beta })]^{-1}$ \cite{LGeAP}.

\section{The Biorthogonal IPR}

The real gauge field does not influence an open system; although the Dirac
norm of eigenstate distribution and the IPR are affected by the asymmetric
coupling, the gauge invariant properties of imaginary field are reflected
from the biorthogonal basis. To better understand the influence of
asymmetric coupling and the imaginary gauge field, we investigate the
biorthogonal norm of the eigenstates. According to the right eigenstates in
Eq. (\ref{ES}), the left eigenstates of the uniform chain satisfies the Schr%
\"{o}dinger equation $H_{\mathrm{chain}}^{\dagger }\left\vert \psi _{n}^{%
\mathrm{L}}\right\rangle =\varepsilon _{n}^{\ast }\left\vert \psi _{n}^{%
\mathrm{L}}\right\rangle $. The left eigenstates are given by $\left\vert
\psi _{n}^{\mathrm{L}}\right\rangle =(\Omega _{n}/\Lambda _{n})\sum_{j=1}^{N}%
\tilde{f}_{j}^{n}a_{j}^{\dagger }\left\vert \mathrm{vac}\right\rangle $ with%
\begin{equation}
\tilde{f}_{j}^{n}=e^{\phi j}\sin \left( k_{n}j\right) ,\Lambda
_{n}=\sum_{j=1}^{N}\left\vert \tilde{f}_{j}^{n}f_{j}^{n}\right\vert ,
\end{equation}%
the left and right eigenstates satisfy the biorthogonal relation%
\begin{equation}
\langle \psi _{m}^{\mathrm{L}}\left\vert \psi _{n}^{\mathrm{R}}\right\rangle
=\delta _{mn},
\end{equation}%
for any integer $m,n\in \lbrack 1,N]$. The biorthogonal norm of the right
eigenstates $\left\vert \psi _{n}^{\mathrm{R}}\right\rangle $ is defined as $%
\langle \psi _{n}^{\mathrm{L}}\left\vert \psi _{n}^{\mathrm{R}}\right\rangle
$. Typical eigenstates distribution under the biorthogonal norm is plotted
in Fig. \ref{fig4}(c), where we observe the extended states in the form of
sinusoidal functions.

Under the biorthogonal norm, the imaginary gauge field has no influence in
the system under open boundary condition; in contrast, under the Dirac norm,
the real gauge field has no influence in the system under open boundary
condition. This reflects the gauge invariant properties of the imaginary
gauge field induced by the asymmetric coupling. The biorthogonal IPR,
defined from both the left and right eigenstates, is given by
\begin{equation}
\tilde{\chi}_{n}=\frac{\sum_{j=1}^{N}\left\vert \tilde{f}_{j}^{n}f_{j}^{n}%
\right\vert ^{2}}{\left( \sum_{j=1}^{N}\left\vert \tilde{f}%
_{j}^{n}f_{j}^{n}\right\vert \right) ^{2}}.
\end{equation}%
Correspondingly the averaged biorthogonal IPR is given by $\tilde{\chi}%
=\left( 1/N\right) \sum_{n=1}^{N}\tilde{\chi}_{n}$. Through a direct
simplification after substituting the expressions of $f_{j}^{n}$ and $\tilde{%
f}_{j}^{n}$, we obtain the averaged biorthogonal IPR as%
\begin{equation}
\tilde{\chi}=\frac{3}{2\left( N+1\right) }.
\end{equation}%
The averaged biorthogonal IPR is inversely proportional to the system size, which implies that the eigenstate is extended and reveals that
the asymmetric coupling induces a gauge field under the biorthogonal norm.
In contrast, the intensity localization under the Dirac norm indicates the
one-way amplification or attenuation. The average IPR efficiently evaluates
the localization of eigenstates. The proposed asymmetric coupling provides a
new route toward nonreciprocal phenomena in photonics, which are valuable
for the unidirectional light modulation and optical isolation.

\section{Conclusion}

In summary, we investigate a uniform non-Hermitian lattice with asymmetric
coupling. The asymmetric coupling induces an imaginary magnetic flux
enclosed in a closed ring configuration, this directly leads to the
non-Hermitian phase transition that all the eigenvalues become complex
except for the special momentum $k=\pi ,2\pi $; however, for system with a
nonuniform coupling, the imaginary magnetic flux does not inevitably induce
the non-Hermitian phase transition. Under open boundary condition, the eigen
values are not influenced by the asymmetric coupling; however, the
eigenstates are localized at the system boundary due to the amplification or
attenuation effect of the asymmetric coupling. The scaling law for the
eigenstates is shown by employing the averaged IPR. The left eigenstates and
biorthogonal norm are studied. The biorthogonal probability distribution of
eigenstates is a sinusoidal function, being extended. The averaged
biorthogonal IPR defined from both left and right eigenstates is inversely
proportional to the system size. Our findings reflect the
underlying physics induced by the asymmetric coupling and are valuable for
further investigations of nonreciprocity and non-Hermitian phase transition
in photonics and beyond.

\section*{Acknowledgement}

We acknowledge the support by National Natural Science Foundation of China
(Grants No. 11605094 and No. 11874225), and the Fundamental Research Funds
for the Central Universities, Nankai University (Grants No. 63191522 and No.
63191738).

\section*{Appendix}

In the Appendix, we show the way of obtaining eigenstate of the system in
detail. The Schr\"{o}dinger equations of the Hamiltonian $H\left\vert \psi
_{n}^{\mathrm{R}}\right\rangle =\sqrt{\alpha \beta }\varepsilon
_{n}\left\vert \psi _{n}^{\mathrm{R}}\right\rangle $\ yield a set of linear
equations%
\begin{equation}
f_{j+1}^{n}-e^{-\phi }\varepsilon _{n}f_{j}^{n}+e^{-2\phi }f_{j-1}^{n}=0,
\end{equation}%
for $2<j<N-1$; at the lattice boundary, we have two other equations%
\begin{equation}
\left\{
\begin{array}{c}
-\varepsilon _{n}f_{1}^{n}+e^{\phi }f_{2}^{n}=0 \\
-\varepsilon _{n}f_{N}^{n}+e^{-\phi }f_{N-1}^{n}=0%
\end{array}%
\right. .
\end{equation}%
We rewrite Schr\"{o}dinger equation $f_{j+1}^{n}-e^{-\phi }\varepsilon
_{n}f_{j}^{n}+e^{-2\phi }f_{j-1}^{n}=0$ as%
\begin{equation}
f_{j+1}^{n}-\mu f_{j}^{n}=\lambda \left( f_{j}^{n}-\mu f_{j-1}^{n}\right) ,
\label{AS}
\end{equation}%
where $j=2,\cdots ,N-1$, and $\left\{ f_{j}^{n}-\mu f_{j-1}^{n}\right\} $\
is ageometric series. Compared the coefficients with the original Schr\"{o}%
dinger equation, we obtain%
\begin{equation}
\left\{
\begin{array}{c}
\mu +\lambda =e^{-\phi }\varepsilon _{n} \\
\lambda \mu =e^{-2\phi }%
\end{array}%
\right. ,  \label{CE}
\end{equation}%
$\mu ,\lambda $ are the roots of the above equations%
\begin{equation}
\mu ,\lambda =\frac{\varepsilon _{n}\pm \sqrt{\varepsilon _{n}^{2}-4}}{%
2e^{\phi }}.  \label{Root}
\end{equation}%
According to Eqs. (\ref{AS}) and (\ref{Root}), we get%
\begin{equation}
\left\{
\begin{array}{c}
f_{j}^{n}-\mu f_{j-1}^{n}=\lambda ^{j-2}\left( f_{2}^{n}-\mu f_{1}^{n}\right)
\\
f_{j}^{n}-\lambda f_{j-1}^{n}=\mu ^{j-2}\left( f_{2}^{n}-\lambda
f_{1}^{n}\right)%
\end{array}%
\right. .  \label{ASset}
\end{equation}%
Substituting the second line into the first line we obtain the expression%
\begin{equation}
f_{j}^{n}=\frac{\left( f_{2}^{n}-\mu f_{1}^{n}\right) \lambda ^{j-1}-\left(
f_{2}^{n}-\lambda f_{1}^{n}\right) \mu ^{j-1}}{\lambda -\mu }.  \label{GF}
\end{equation}%
Assuming $f_{1}^{n}=1$, based on one boundary condition $\varepsilon
_{n}e^{-\phi }f_{1}^{n}=f_{2}^{n}$ and the first line of Eq. (\ref{CE}), Eq.
(\ref{GF}) reduces to
\begin{equation}
f_{j}^{n}=\frac{\lambda ^{j}-\mu ^{j}}{\lambda -\mu }.
\end{equation}%
After substituting $f_{j}^{n}$\ into anther boundary condition $\varepsilon
_{n}f_{N}^{n}=e^{-\phi }f_{N-1}^{n}$, employing Eq. (\ref{CE}), we obtain%
\textbf{\ }%
\begin{equation}
\left( \lambda e^{\phi }\right) ^{2\left( N+1\right) }=1,
\end{equation}%
which means%
\begin{equation}
\lambda e^{\phi }=e^{ik_{n}},
\end{equation}%
for $k_{n}=n\pi /\left( N+1\right) $ with $n=1,\cdots ,N$. From Eq. (\ref%
{Root}), we obtain $\varepsilon _{n}+\sqrt{\varepsilon _{n}^{2}-4}%
=2e^{ik_{n}}$ or $\varepsilon _{n}-\sqrt{\varepsilon _{n}^{2}-4}=2e^{ik_{n}}$%
. In any case, the energy is given by
\begin{equation}
\varepsilon _{n}=2\cos k_{n}.
\end{equation}%
Substituting $\varepsilon _{n}$ into $f_{j}^{n}$ and considering the
normalization factor, we obtain the wave functions%
\begin{equation}
f_{j}^{n}=e^{-\phi j}\sin \left( k_{n}j\right) ,
\end{equation}%
where $j=1,\cdots ,N$. The Dirac normalization factor is
\begin{equation}
\Omega _{n}=\sqrt{\sum_{j=1}^{N}e^{-2j\phi }\sin ^{2}\left( k_{n}j\right) }.
\end{equation}


\begin{thebibliography}{99}
\bibitem{Bender} C. M. Bender, Rep. Prog. Phys. \textbf{70}, 947 (2007).

\bibitem{NM} N. Moiseyev, Non-Hermitian Quantum Mechanics (Cambridge Univ.
Press, 2011).

\bibitem{NatPhoton} L. Feng, R. El-Ganainy, and L. Ge, Nat. Photo. \textbf{11%
}, 752 (2017).

\bibitem{SLonghi} S. Longhi, Eur. Phys. Lett. \textbf{120}, 64001 (2017).

\bibitem{Fan} Y. Huang, Y. Shen, C. Min, S. Fan, and G. Veronis,
Nanophotonics \textbf{6}, 977 (2017).

\bibitem{El-G} R. El-Ganainy, K. G. Makris, M. Khajavikhan, Z. H.
Musslimani, S. Rotter, and D. N. Christodoulides,\ Nat. Phys. \textbf{14},
11 (2018).

\bibitem{SKG} S. K. Gupta, Y. Zou, X. Y. Zhu, M. H. Lu, L. Zhang, X. P. Liu,
and Y. F. Chen, arXiv:1803.00794.

\bibitem{Alu2019} M. A. Miri and A. Al\`{u}, Science \textbf{363}, eaar7709
(2019).

\bibitem{AGuo} A. Guo, G. J. Salamo, D. Duchesne, R. Morandotti, M.
Volatier-Ravat, V. Aimez, G. A. Siviloglou, and D. N. Christodoulides, Phys.
Rev. Lett. \textbf{103}, 093902 (2009).

\bibitem{EP} C. Dembowski, B. Dietz, H. D. Gr\"{a}f, H. L. Harney, A. Heine,
W. D. Heiss, and A. Richter, Phys. Rev. E \textbf{69}, 056216 (2004); M. M%
\"{u}ler and I. Rotter, J. Phys. A \textbf{41}, 244018 (2008); R. Uzdin, A.
Mailybaev, and N. Moiseyev, J. Phys. A \textbf{44}, 435302 (2011).

\bibitem{EP2} W. D. Heiss, J. Phys. A \textbf{45}, 444016 (2012).

\bibitem{BZhen} B. Zhen, C. W. Hsu, Y. Igarashi, L. Lu, I. Kaminer, A. Pick,
S. L. Chua, J. D. Joannopoulos, and M. Solja\v{c}i\'{c}, Nature (London)
\textbf{525}, 354 (2015).

\bibitem{Doppler} J. Doppler, A. A. Mailybaev, J. B\"{o}hm, U. Kuhl, A.
Girschik, F. Libisch, T. J. Milburn, P. Rabl, N. Moiseyev, and S. Rotter,
Nature (London) \textbf{537}, 76 (2016).

\bibitem{HXu} H. Xu, D. Mason, L. Jiang, and J. G. E. Harris, Nature
(London) \textbf{537}, 80 (2016).

\bibitem{CTChanPRX} K. Ding, G. Ma, M. Xiao, Z. Q. Zhang, and C. T. Chan,
Phys. Rev. X \textbf{6}, 021007 (2016); X. L. Zhang, S. B. Wang, B. Hou, and
C.\thinspace T. Chan, Phys. Rev. X \textbf{8}, 021066 (2018).

\bibitem{Assawaworrarit} S. Assawaworrarit, X. Yu, and S. Fan, Nature
\textbf{546}, 387 (2017).

\bibitem{HJing2017} H. Jing, S. K. Ozdemir, H. Lu, and F. Nori, Sci. Rep.
\textbf{7}, 3386 (2017).

\bibitem{JL97} L. Jin, Phys. Rev. A \textbf{97}, 012121 (2018).

\bibitem{XLCui} L. Pan, S. Chen, and X. Cui, Phys. Rev. A \textbf{99},
011601(R) (2019).

\bibitem{WP} P. Wang, L. Jin, G. Zhang, and Z. Song, Phys. Rev. A \textbf{94}%
, 053834 (2016); Q. Zhong, D. N. Christodoulides, M. Khajavikhan, K. G.
Makris, and R. El-Ganainy, Phys. Rev. A \textbf{97}, 020105(R) (2018); L.
Ge, Photon. Res. \textbf{6}, A10 (2018).

\bibitem{NMParaEstimation} M. Am-Shallem, R. Kosloff, and N. Moiseyev, New
J. Phys. \textbf{17}, 113036 (2015).

\bibitem{JW} J. Wiersig, Phys. Rev. Lett. \textbf{112}, 203901 (2014); W.
Chen, S. K. Ozdemir, G. Zhao, J. Wiersig, and L. Yang, Nature (London)
\textbf{548}, 192 (2017); H. Hodaei, A. U. Hassan, S. Wittek, H.
Garcia-Gracia, R. El-Ganainy, D. N. Christodoulides, and M. Khajavikhan,
Nature (London) \textbf{548}, 187 (2017).

\bibitem{YXLiu} Z. P. Liu, J. Zhang, \c{S}. K. \"{O}demir, B. Peng, H. Jing,
X. Y. L\"{u}, C. W. Li, L. Yang, F. Nori, and Y. X. Liu, Phys. Rev. Lett.
\textbf{117}, 110802 (2016).

\bibitem{Makris08} K. G. Makris, R. El-Ganainy, D. N. Christodoulides, and
Z. H. Musslimani, Phys. Rev. Lett. \textbf{100}, 103904 (2008).

\bibitem{Klaiman} S. Klaiman, U. Gunther, and N. Moiseyev, Phys. Rev. Lett.
\textbf{101}, 080402 (2008).

\bibitem{Ruter} C. E. R\"{u}ter, K. G. Makris, R. El-Ganainy, D. N.
Christodoulides, M. Segev, and D. Kip, Nat. Phys. \textbf{6}, 192 (2010).

\bibitem{YDChongCPA} Y. D. Chong, L. Ge, H. Cao, and A. D. Stone, Phys. Rev.
Lett. \textbf{105}, 053901 (2010); W. Wan, Y. Chong, L. Ge, H. Noh, A. D.
Stone, and H. Cao, Science \textbf{331}, 889 (2011).

\bibitem{Nature2012} A. Regensburger, C. Bersch, M. A. Miri, G. Onishchukov,
D. N. Christodoulides, and U. Peschel, Nature (London), \textbf{488}, 167
(2012).

\bibitem{LFengNatureMater} L. Feng, Y. L. Xu, W. S. Fegadolli, M. H. Lu, Jos%
\'{e}. E. B. Oliveira, V. R. Almeida, Y. F. Chen, and A. Scherer, Nat.
Mater. \textbf{12}, 108 (2013).

\bibitem{UPA} H. Ramezani, Y. Wang, E. Yablonovitch, and X. Zhang, Ieee. J.
Sel. Top. Quant. \textbf{22}, 5000706\ (2016); L. Jin, P. Wang, and Z. Song,
Sci. Rep. \textbf{6}, 32919 (2016);

\bibitem{LJin} L. Jin and Z. Song, Phys. Rev. Lett. \textbf{121}, 073901
(2018).

\bibitem{XZhang} H. Ramezani, H. K. Li, Y. Wang, and X. Zhang, Phys. Rev.
Lett. \textbf{113}, 263905 (2014).

\bibitem{LYangPNAS} B. Peng, S. K. \"{O}zdemir, M. Liertzer, W. Chen, J.
Kramer, H. Y\i lmaz, J. Wiersig, S. Rotter, and L. Yang, Proc. Natl. Acad.
Sci. U.S.A. \textbf{113}, 6845 (2016).

\bibitem{Kominis16} Y. Kominis, T. Bountis, and S. Flach, Sci. Rep. \textbf{6%
}, 33699 (2016); Phys. Rev. A \textbf{95}, 063832 (2017); Y. Kominis, K. D.
Choquette, T. Bountis, and V. Kovanis, Appl. Phys. Lett. \textbf{113},
081103 (2018).

\bibitem{Wu} J. H. Wu, M. Artoni, and G. C. La Rocca, Phys. Rev. Lett.
\textbf{113}, 123004 (2014); Z. Zhang, Y. Zhang, J. Sheng, L. Yang, M. A.
Miri, D. N. Christodoulides, B. He, Y. Zhang, and M. Xiao, Phys. Rev. Lett.
\textbf{117}, 123601 (2016); P. Peng, W. Cao, C. Shen, W. Qu, J. Wen, L.
Jiang, and Y. Xiao, Nat. Phys. \textbf{12}, 1139 (2016).

\bibitem{Ruschhaupt} A. Ruschhaupt, F. Delgado, and J. G. Muga, J. Phys. A
\textbf{38}, L171 (2005).

\bibitem{GanainyOL} R. El-Ganainy, K. G. Makris, D. N. Christodoulides, and
Z. H. Musslimani, Opt. Lett. \textbf{32}, 2632 (2007).

\bibitem{YFChen16} Y. L. Xu, W. S. Fegadolli, L. Gan, M. H. Lu, X. P. Liu,
Z. Y. Li, A. Scherer, and Y. F. Chen, Nat. Commun. \textbf{7}, 11319 (2016).

\bibitem{BPeng} B. Peng, S. K. \"{O}zdemir, F. Lei, F. Monifi, M. Gianfreda,
G. L. Long, S. Fan, F. Nori, C. M. Bender, and L. Yang, Nat. Phys. \textbf{10%
}, 394 (2014).

\bibitem{HJing} H. Jing, S. K. Ozdemir, X. Y. Lu, J. Zhang, L. Yang, and F.
Nori, Phys. Rev. Lett. \textbf{113}, 053604 (2014).

\bibitem{He} B. He, L. Yang, and M. Xiao, Phys. Rev. A \textbf{94},
031802(R) (2016); B. He, L. Yang, X. Jiang, and M. Xiao, Phys. Rev. Lett.
\textbf{120}, 203904 (2018).

\bibitem{Cerjan} A. Cerjan, A. Raman, and S. Fan, Phys. Rev. Lett. \textbf{%
116}, 203902 (2016).

\bibitem{YFChenPNAS} C. He, X. C. Sun, X. P. Liu, M. H. Lu, Y. Chen, L.
Feng, and Y. F. Chen, Proc. Natl. Acad. Sci. U.S.A. \textbf{113}, 4924
(2016).

\bibitem{Kottos} J. Schindler, A. Li, M. C. Zheng, F. M. Ellis, and T.
Kottos, Phys. Rev. A \textbf{84}, 040101(R) (2011); J. Schindler, Z. Lin, J.
M. Lee, H. Ramezani, F. M. Ellis, and T. Kottos, J. Phys. A: Math. Theor.
\textbf{45}, 444029 (2012).

\bibitem{Alu} R. Fleury, D. Sounas, and A. Al\`{u}, Nat. Commun. \textbf{6},
5905 (2015); S. A. Cummer, J. Christensen, and A. Al\`{u}, Nat. Rev. Mater.
\textbf{1}, 16001 (2016).

\bibitem{XZhangPRX} X. Zhu, H. Ramezani, C. Shi, J. Zhu, and X. Zhang, Phys.
Rev. X \textbf{4}, 031042 (2014).

\bibitem{LChang} L. Chang, X. Jiang, S. Hua, C. Yang, J. Wen, L. Jiang, G.
Li, G. Wang, and M. Xiao, Nat. Photon. \textbf{8}, 524 (2014).

\bibitem{XZhangScience} L. Feng, Z. J. Wong, R. M. Ma, Y. Wang, and X.
Zhang, Science \textbf{346}, 972 (2014).

\bibitem{PNAS} B. Peng, S. K. Ozdemir, M. Liertzer, W. J. Chen, J. Kramer,
H. Yilmaz, J. Wiersig, S. Rotter, and L. Yang, Proc. Natl. Acad. Sci.
U.S.A., \textbf{113}, 6845 (2016).

\bibitem{TopoLasing} G. Harari, M. A. Bandres, Y. Lumer, M. C. Rechtsman, Y.
D. Chong, M. Khajavikhan, D. N. Christodoulides, and M. Segev, Science
\textbf{359}, eaar4003 (2018); M. A. Bandres, S. Wittek, G. Harari, M.
Parto, J. Ren, M. Segev, D. Christodoulides, and M. Khajavikhan, Science
\textbf{359}, eaar4005 (2018).

\bibitem{NatPhysAntiPT} P. Peng, W. Cao, C. Shen, W. Qu, J. Wen, L. Jiang,
and Y. Xiao, Nat. Phys. \textbf{12}, 1139 (2016).

\bibitem{Longhi} S. Longhi, D. Gatti, and G. Della Valle, Sci. Rep. \textbf{5%
}, 13376 (2015); Phys. Rev. B \textbf{92}, 094204 (2015).

\bibitem{LFengNC} B. Midya, H. Zhao, and L. Feng, Nat. Commun. \textbf{9},
2674 (2018).

\bibitem{ZGong} Z. P. Gong, Y. Ashida, K. Kawabata, K. Takasan, S.
Higashikawa, and M. Ueda, Phys. Rev. X \textbf{8}, 031079 (2018).

\bibitem{JL} L. Jin, Phys. Rev. A \textbf{96}, 032103 (2017); L. Jin, P.
Wang, and Z. Song, Sci. Rep. \textbf{7}, 5903 (2017).

\bibitem{Weimann} S. Weimann, M. Kremer, Y. Plotnik, Y. Lumer, S. Nolte, K.
G. Makris, M. Segev, M. C. Rechtsman, and A. Szameit, Nat. Mater. \textbf{16}%
, 433 (2017).

\bibitem{Zeuner} J. M. Zeuner, M. C. Rechtsman, Y. Plotnik, Y. Lumer, S.
Nolte, M. S. Rudner, M. Segev, and A. Szameit, Phys. Rev. Lett. \textbf{115}%
, 040402 (2015).

\bibitem{Schomerus} H. Schomerus, Opt. Lett. \textbf{38}, 1912 (2013); S.
Malzard, C. Poli, and H. Schomerus, Phys. Rev. Lett. \textbf{115}, 200402
(2015); C. Poli, M. Bellec, U. Kuhl, F. Mortessagne, and H. Schomerus, Nat.
Commun. \textbf{6}, 6710 (2015).

\bibitem{HZhao} H. Zhao, S. Longhi, and L. Feng, Sci. Rep. \textbf{5}, 17022
(2015); M. Pan, H. Zhao, P. Miao, S. Longhi, and L. Feng, Nat. Commun.
\textbf{9}, 1308 (2018); B. Midya and L. Feng, Phys. Rev A 98, 043838 (2018).

\bibitem{HZhao2} H. Zhao, P. Miao, M. H. Teimourpour, S. Malzard, R.
El-Ganainy, H. Schomerus, and L. Feng, Nat. Commun. \textbf{9}, 981 (2018);
M. Parto, S. Wittek, H. Hodaei, G. Harari, M. A. Bandres, J. Ren, M. C.
Rechtsman, M. Segev, D. N. Christodoulides, and M. Khajavikhan, Phys. Rev.
Lett. \textbf{120}, 113901 (2018).

\bibitem{2013} G. Q. Liang and Y. D. Chong, Phys. Rev. Lett. \textbf{110},
203904 (2013); B. Zhu, R. L\"{u}, and S. Chen, Phys. Rev. A \textbf{89},
062102 (2014); C. Yin, H. Jiang, L. Li, R. L\"{u}, and S. Chen, Phys. Rev. A
\textbf{97}, 052115 (2018); H. Jiang, C. Yang, and S. Chen, Phys. Rev. A
\textbf{98}, 052116 (2018); C. H. Liu, H. Jiang, S. Chen, arXiv:1812.04819.

\bibitem{2011} S. Diehl, E. Rico, M. A. Baranov, and P. Zoller, Nat. Phys.
\textbf{7}, 971 (2011); Y. C. Hu and T. L. Hughes, Phys. Rev. B \textbf{84},
153101 (2011); K. Esaki, M. Sato, K. Hasebe, and M. Kohmoto, Phys. Rev. B
\textbf{84}, 205128 (2011).

\bibitem{JGong15} J. Gong and Q. H. Wang, Phys. Rev. A \textbf{91}, 042135
(2015); L. Zhou, Q. H. Wang, H. Wang, and J. Gong, Phys. Rev. A \textbf{98},
022129 (2018).

\bibitem{Leykam} D. Leykam, K. Y. Bliokh, C. Huang, Y. D. Chong, and F.
Nori, Phys. Rev. Lett. \textbf{118}, 040401 (2017); T. Liu, Y. R. Zhang, Q.
Ai, Z. Gong, K. Kawabata, M. Ueda, and F. Nori, Phys. Rev. Lett. \textbf{122}%
, 076801 (2018).

\bibitem{YXu} Y. Xu, S. T. Wang, and L. M. Duan, Phys. Rev. Lett. \textbf{118%
}, 045701 (2017); Q. B. Zeng, Y. B. Yang, and Y. Xu, arXiv:1901.08060.

\bibitem{HShen} H. Shen, B. Zhen, and L. Fu, Phys. Rev. Lett. \textbf{120},
146402 (2018); H. Shen and L. Fu, Phys. Rev. Lett. \textbf{121}, 026403
(2018).

\bibitem{Yoko} K. Yokomizo and S. Murakami, arXiv:1902.10958.

\bibitem{Yuce} C. Yuce, Phys. Lett. A \textbf{379}, 1213 (2015); Phys. Rev.
A \textbf{93}, 062130 (2016); Phys. Rev. A \textbf{97}, 042118 (2018); Phys.
Rev. A \textbf{98}, 012111 (2018); Z. Oztas and C. Yuce, Phys. Rev. A
\textbf{98}, 042104 (2018); Z. Ozcakmakli Turker and C. Yuce, Phys. Rev. A
\textbf{99}, 022127 (2019).

\bibitem{Lieu} S. Lieu, Phys. Rev. B \textbf{97}, 045106 (2018); Phys. Rev.
B \textbf{98}, 115135 (2018).

\bibitem{Molina} F. Munoz, F. Pinilla, J. Mella, and M. I. Molina, Sci. Rep.
\textbf{8}, 17330 (2018).

\bibitem{Kunst} F. K. Kunst, E. Edvardsson, J. C. Budich, and E. J.
Bergholtz, Phys. Rev. Lett. \textbf{121}, 026808 (2018); J. Carlstr\"{o}m
and E. J. Bergholtz, Physical Review A \textbf{98}, 042114 (2018); E.
Edvardsson, F. K. Kunst, and E. J. Bergholtz, arXiv:1812.09060; J. Carlstr%
\"{o}m, M. St\aa lhammar, J. C. Budich, and E. J. Bergholtz,
arXiv:1810.12314; J. C. Budich, J. Carlstr\"{o}m, F. K. Kunst, and E. J.
Bergholtz, Phys. Rev. B \textbf{99}, 041406(R) (2019).

\bibitem{KK} K. Kawabata, K. Shiozaki, and M. Ueda, Phys. Rev. B \textbf{98}%
, 165148 (2018); K. Kawabata, Y. Ashida, H. Katsura, and M. Ueda, Phys. Rev.
B \textbf{98}, 085116 (2018); K. Kawabata, S. Higashikawa, Z. Gong, Y.
Ashida, and M. Ueda, Nat. Commun. \textbf{10}, 297 (2019); K. Kawabata, K.
Shiozaki, M. Ueda, and M. Sato, arXiv:1812.09133.

\bibitem{LJLPRB} L. J. Lang, Y. Wang, H. Wang, and Y. D. Chong, Phys. Rev. B
\textbf{98}, 094307 (2018); X. Ni, D. Smirnova, A. Poddubny, D. Leykam, Y.
Chong, and A. B. Khanikaev, Phys. Rev. B \textbf{98}, 165129 (2018).

\bibitem{JHou} J. Hou, Z. Li, X. W. Luo, Q. Gu, and C. Zhang,
arXiv:1808.06972.

\bibitem{Takata} K. Takata and M. Notomi, Phys. Rev. Lett. \textbf{121},
213902 (2018).

\bibitem{Okugawa} R. Okugawa and T. Yokoyama, Phys. Rev. B \textbf{99},
041202(R) (2019).

\bibitem{TDas} A. Ghatak and T. Das, Phys. Rev. B \textbf{97}, 014512
(2018); B. X. Wang and C. Y. Zhao, Phy. Rev. B \textbf{98}, 165435 (2018);
Phy. Rev. A \textbf{98}, 023808 (2018).

\bibitem{VMMA} K. Moors, A. A. Zyuzin, A. Y. Zyuzin, R. P. Tiwari, and T. L.
Schmidt, Phy. Rev. B \textbf{99}, 041116(R) (2019); V. M. Martinez Alvarez,
J. E. Barrios Vargas, and L. E. F. Foa Torres, Phys. Rev. B \textbf{97},
121401(R) (2018); V. M. Martinez Alvarez and M. D. Coutinho-Filho, Phys.
Rev. A \textbf{99}, 013833 (2019).

\bibitem{HZhou} H. Zhou and J. Y. Lee, arXiv: 1812.10490; F. K. Kunst and V.
Dwivedi, arXiv:1812.02186.

\bibitem{Ezawa} M. Ezawa, arXiv:1810.04527; arXiv:1811.12059;
arXiv:1902.03716; L. Herviou, J. H. Bardarson, and N. Regnault,
arXiv:1901.0001; H. G. Zirnstein, G. Refael, and B. Rosenow,
arXiv:1901.11241; M. R. Hirsbrunner, T. M. Philip, and M. J. Gilbert,
arXiv:1901.09961;

\bibitem{Dan} D. S. Borgnia, A. J. Kruchkov, and R. J. Slager,
arXiv:1902.07217.

\bibitem{Torres} V. M. Martinez Alvarez, J. E. Barrios Vargas, M. Berdakin,
and L. E. F. Foa Torres, Eur. Phys. J. Spec. Top. \textbf{227}, 1295 (2018).

\bibitem{ZZL} Z. Z. Li, X. S. Li, L. L. Zhang, and W. J. Gong,
arXiv:1901.10688.

\bibitem{ZWang} S. Yao and Z. Wang, Phys. Rev. Lett. \textbf{121}, 086803
(2018); S. Yao, F. Song, and Z. Wang, Phys. Rev. Lett. \textbf{121}, 136802
(2018).

\bibitem{ZYang} Z. Yang and J. Hu, Phys. Rev. B \textbf{99}, 081102(R)
(2019).

\bibitem{LJinPRB} L. Jin and Z. Song, Phys. Rev. B \textbf{99}, 081103(R)
(2019).

\bibitem{HWang} H. Wang, J. Ruan, and H. Zhang, Phys. Rev. B \textbf{99},
075130 (2019).

\bibitem{CHLee} C. H. Lee and R. Thomale, arXiv:1809.02125; C. H. Lee, L.
Li, and J. Gong, arXiv:1810.11824; C. H. Lee, G. Li, Y. Liu, T. Tai, R.
Thomale, and X. Zhang, arXiv: 1812.02011.

\bibitem{RYu} K. Luo, J. Feng, Y. X. Zhao, and R. Yu, arXiv:1810.09231

\bibitem{HJiang2019} H. Jiang, L. J. Lang, C. Yang, S. L. Zhu, and S. Chen,
arXiv:1901.09399.

\bibitem{DasReview} A. Ghatak and T. Das, arXiv:1902.07972.

\bibitem{Degree} D. D. Scott and Y. N. Joglekar, Phys. Rev. A \textbf{83},
050102(R) (2011).

\bibitem{LonghiLPR} S. Longhi, Laser \& Photo. Rev. \textbf{3}, 243 (2009).

\bibitem{LGeAP} L. Ge, Ann. Phys. (Berlin) \textbf{529}, 1600182 (2017).
\end{thebibliography}
\end{document}